\documentclass[letterpaper]{article}
\usepackage{iccc}

\usepackage{times}
\usepackage{helvet}
\usepackage{courier}
\usepackage{natbib}

\usepackage{url}
\usepackage{hyperref}
\usepackage{subcaption}
\usepackage{graphicx}
\usepackage{amsmath}
\usepackage{algorithm}
\usepackage[noend]{algpseudocode}

\title{Performing Structured Improvisations with pre-trained Deep Learning Models}
\author{
  Pablo Samuel Castro\\
  Google Brain\\
  psc@google.com\\
}
\begin{document} 
\maketitle
\begin{abstract}
  The quality of outputs produced by deep generative models for music
  have seen a dramatic improvement in the last few years. However, most deep
  learning models perform in ``offline'' mode, with few restrictions on the
  processing time.  Integrating these types of models into a live structured
  performance poses a challenge because of the necessity to respect the beat
  and harmony. Further, these deep models tend to be agnostic to the style of
  a performer, which often renders them impractical for live performance.
  In this paper we propose a system which enables the integration of
  out-of-the-box generative models by leveraging the musician’s creativity and
  expertise.
\end{abstract}

\section{Introduction}
The popularity and quality of machine learning models has seen a tremendous
growth over the last few years.  {\em Generative models}, which are trained to
produce outputs resembling a pre-specified data distribution, have attracted
much attention from both the scientific and artistic community in large part
due to the realism of the outputs produced. Of note are some of the Generative
Adversarial Networks \citep{goodfellow14gan} which have ben able to yield
high-quality images \citep{brock18large}, image-to-image translations
\citep{zhu17unpaired,wang18pix2pixHD}, and even transfer of a professional dance
performance onto an amateur dancer \citep{chan18dance}. The ability of these
models to produce completely new outputs that are highly realistic has drawn a
growing number of artists to incorporate them into their craft, taking steps
into a new creative frontier.

In the musical domain, recent works produce music that is both realistic and
interpolatable \citep{roberts18hierarchical}, closely resembles human
performance \citep{huang19improved}, and can aid in automatic
composition\footnote{https://www.ampermusic.com/}. The increased realism of
these models is typically accompanied with an increase in the amount of
processing time required to generate outputs. If these outputs can be produced
in an ``offline'' manner (i.e. generated well before they will be consumed) the
longer processing times are not problematic. On the other hand, long processing
times generally renders these models inadequate for live performance.  This
issue is particularly stark in structured improvisation, such as in traditional
jazz, where the music produced must respect the beat and harmony of the piece.

Despite the increased realism and consistency of the musical models mentioned
above, the models are inherently {\em impersonal}: there is a single model for
all users. Artists wishing to incorporate these models into their work are thus
forced to either train their own model -- requiring a good understanding
of machine learning -- or adapt their craft to the idiosyncracies of the model
-- which can result in creative output that is not reflective of the artist's
style.

In this paper we introduce a software system that enables the incorporation of
generative musical models into musical improvisation. This can be used as both
a solo-performance or in an ensemble. Our system produces a performance that is
a hybrid of human improvisation with melodies and rhythms generated by deep 
learning models. As we will describe below, our hybrid approach enables us to
address the two aforementioned problems: real-time compatibility and
stylistic personalization. It is worth specifying that our system does not
require machine learning expertise.

The paper is structured as follows.  We begin by providing some necessary
background on recurrent neural networks and how they are used for rhythm and
melody generation, followed by a discussion of related work incorporating
machine learning models with artistic creation.  We then present our system
setup and describe the software backend in.  Finally, we provide some empirical
evidence of the efficacy of our system, followed by some concluding remarks and
discussion of future avenues of research in.

\begin{figure*}[!t]
  \centering
  \includegraphics[width=0.7\textwidth]{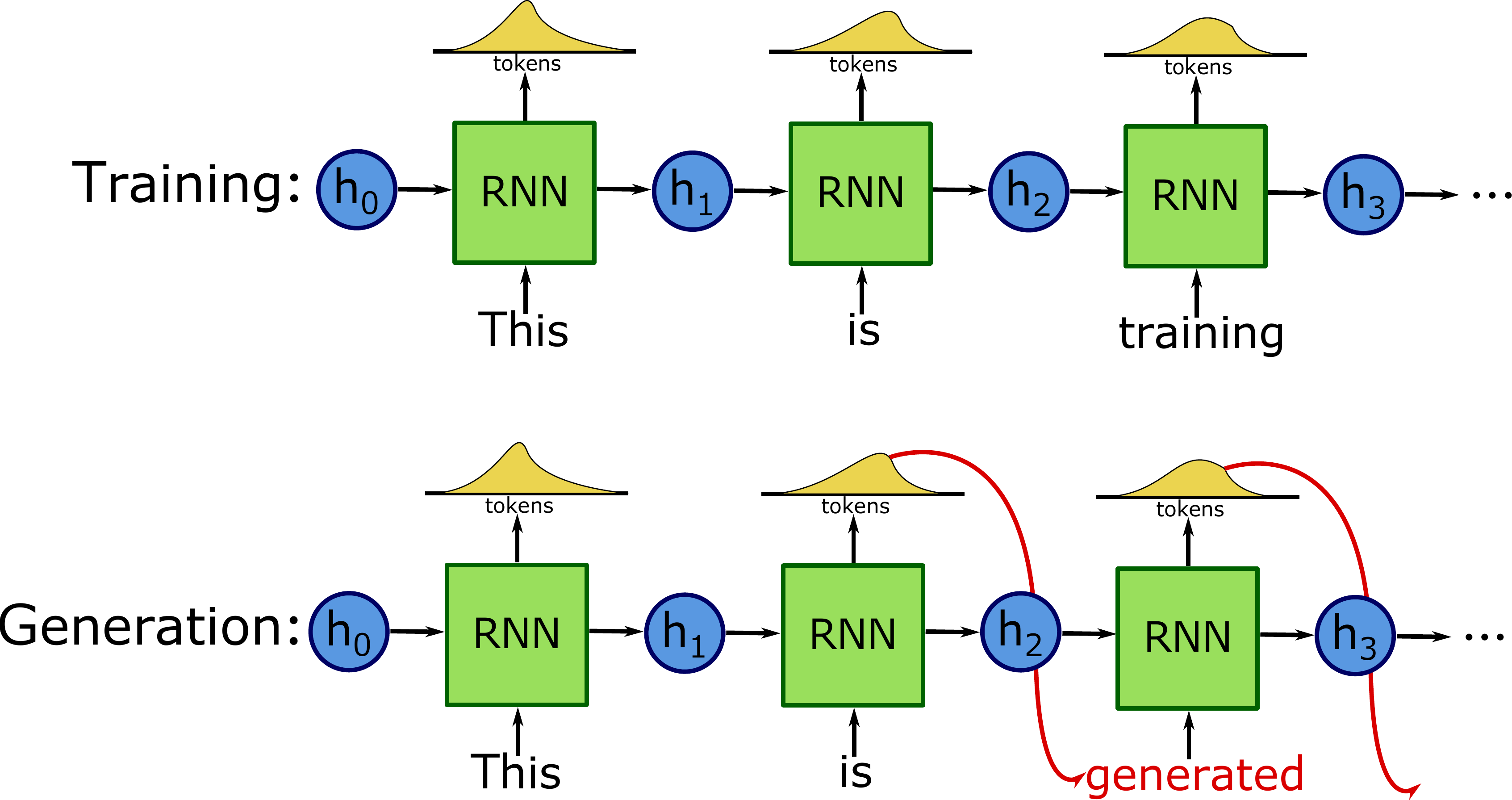}
  \caption{An RNN being trained on english sentences. The green squares
  represent the RNN, where we are explicitly illutrating the updating internal
  states $h_i$ as blue circles. The output of the model is a distribution over
  tokens (in yellow). The top frame shows an example of a training pass through
  the sentence ``This is training ...''.  The bottom frame shows how the model
  can be used for generation when fed in the primer sequence ``This is''. In
  this example, the sample from the model's distribution over tokens is
  `generated'.}
  \label{fig:lstm}
\end{figure*}

\section{Background}
\label{sec:background}
We use Recurrent Neural Networks (RNNs) \citep{rumelhart86learning} as the
machine learning models for generating drum beats and melodies. Recurrent
Neural Networks are a special type of neural network which process a {\em
sequence} of tokenized inputs one token at a time, updating an internal state
after processing each input. This internal state can enable a type of
``memory'' for long-term dependencies. The network maintains a probability
distribution over a finite dictionary of tokens; the model's internal parameters
determine the shape of the distribution conditional on its current internal
state and the last token. During training, the internal model parameters are
updated so as to maximize the probability mass assigned by the network to the
true token $t_{n+1}$, given the previous tokens $t_{1:n}$.  One of the most
popular types of RNN are the Long Short-Term Memory networks introduced by
\cite{hochreiter97lstm}.

A trained RNN can be used for generation: after processing a sequence of tokens
$t_{1:n}$, sample from the resulting internal distribution over the token
dictionary. A properly trained model will produce outputs that have a high
likelihood to have come from the distribution of the training dataset. When
using these models for generation, we will refer to the initial sequence
$t_{1:n}$ fed into the model as the {\em primer sequence}. \autoref{fig:lstm}
illustrates this process for an RNN where the tokens are english words.

We will make use of two LSTM-models from Google Magenta. The first is MelodyRNN
\citep{melodyrnn}.  It processes {\em note events} as tokens, where a note event
contains a note's pitch and its duration. The model assumes monophonic melodies
(i.e. only one note played at a time) and is instrument agnostic. Thousands of
MIDI files were used for training. These MIDI files were quantized into 16th
notes: that is, the minimum allowable time between two notes are one 16th
note\footnote{There are sixteen 16th notes in one bar of 4/4 time.}. The
Magenta team provides a few models trained with different configurations, but
we use the {\em Attention} configuration in this work, as we found it produces
the best results. When using this model to generate new melodies, the melodies
produced tend to match the key signature and note density of the primer melody
sequence fed into it, which is a desirable property for our use case.

The second is DrumsRNN \citep{drumsrnn}.  The model is similar to MelodyRNN, but
here there is polyphony as multiple drums can be hit simultaneously. As for
MelodyRNN, this model was trained on thousands of MIDI files, quantized into
16th notes.

\section{Related Work}
\label{sec:related}
The use of deep learning models for creative purposes has witnessed a steady
increase in the artistic and scientific community in the past few years.
Arguably the most common approach is to train a custom model on a particular
dataset to produce outputs which are novel, but match the distribution of the
training set. A representative example in the visual domain is the Pix2PixHD
model \citep{wang18pix2pixHD} trained on a corpus of Balenciaga runway shows to
produce new ``imagined'' outfits \citep{barrat18balenciaga}.
\cite{mathewson17improvised} and \cite{mathewson19shaping} explore methods for integrating
generative language models with improvised theatre.

In the musical domain, there is a rather stark difference when comparing the
outputs of deep-learning models against ``shallow'' models. We begin our survey
of related work with those that came before the deep-learning revolution, and
then proceed to some works using deep models.

There have been a number of works proposing new types of digital instruments
which make use of machine learning models.  The Wekinator
\citep{fiebrink09wekinator} enables users to train new models in a {\em
supervised} fashion by providing pairs of inputs and expected outputs; inputs
can be provided in many forms including using computer controllers and physical
gestures, while outputs can be sent to any musical, digital or physical
actuator. This contrasts with our proposed framework, which does not
require retraining a model, but rather adapt the outputs of a pre-trained deep
learning model to a performer's style.

\cite{thom00unsupervised} and \cite{thom01machine}
build probabilistic models to emulate an improviser's tonal and melodic trends.
\cite{johnson17learning} makes use of two LSTMs:
one for intervals between notes and the other for note
intervals relative to the underlying chord progression; these trained models
are then combined to generate melodies in a recurrent note-by-note fashion.

In \citep{weinberg09interactive} the authors introduce {\em shimon}, a robot marimba
player capable of interacting with human players. The robot has human-like
movements (such as head-bobbing, ``gazing'' to pass on the solo to another
player, etc.) which make it natural to interact with.

Closely related to our use of `continuations' are {\em The Continuator} of
\cite{pachet03continuator}, where the authors use Markov models to adapt to a
user's style. In contrast to our work, however, the continuator is agnostic to
the underlying beat of a performance, which is essential to jazz improvisation.

\cite{bretan17deep} propose training a deep autoencoder to
encode melodies played by a performer into a latent space that has been trained
to capture musical consistency; the closest melody from a library that has been
embedded into the same latent space is returned, allowing their system to
respond in near real-time. Although the authors augment the original library 
(a large corpus of MIDI files) by ``humanizing'' the samples via application of
temporal shifts, the outputs are still limited to what is available in the
library. Thus, the resulting {\em style} is more reflective of the augmented
library than of the human performer.

\cite{roberts18hierarchical} propose a deep autoencoder model for encoding
melodies into a latent space, combined with a deep decoder for converting
points from that latent space into cohesive melodies.

\cite{huang19improved} trained a transformer model \citep{vaswani17attention} on
a dataset of virtuoso piano performances, resulting in a model that can produce
highly realistic and novel musical snippets\footnote{See
https://magenta.tensorflow.org/music-transformer for some examples.}.

More recently, a number of creative web applications combine some of the
deep-learning models mentioned above with interactive interfaces that allow
users with no musical training to create music
\citep{parviainen18incredible,dinculescu18tenori}.  \cite{hantrakul18gesture}
use an RNN to convert gestures on a Roli Lightpad into sound.

These digital instruments, however, tend to operate in isolation; specifically,
they prescribe the beat and harmony (or are agnostic to it), rather than
conforming to pre-existing beats and harmonies. In jazz performance, especially
when performing with other musicians, respecting these is crucial
when improvising\footnote{We are excluding free jazz here, which allows for
more flexibility in performance.}. In this work we propose a system enabling
the integration of out-of-the-box deep generative models for this type of
performance.

\section{System setup}
\label{sec:system}
\begin{figure*}[!h]
  \centering
  \includegraphics[width=0.7\textwidth]{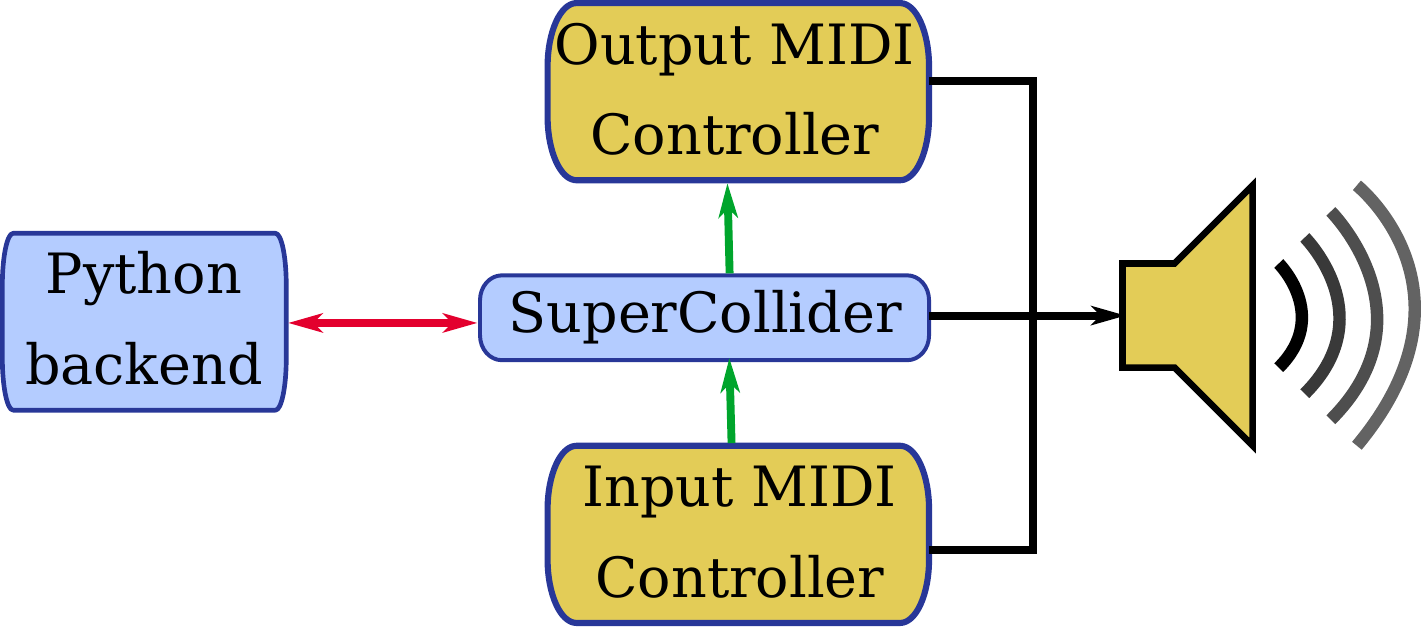}
  \caption{The main components of our system. Orange and blue boxes indicate
  hardware and software components, respectively. Black arrows indicate raw
  audio signals, green arrows indicate MIDI signals, and red arrows indicate
  OSC signals.}
  \label{fig:setup}
\end{figure*}

Our setup assumes a piano keyboard connected to a computer via MIDI used for
input, along with an additional controller for enabling more MIDI control
messages; in our case we are using the Korg Nanokontrol2 MIDI controller but
the system can be used with any MIDI controller.  We use
SuperCollider\footnote{https://supercollider.github.io/} to detect all incoming
MIDI events and pipe them as OSC\footnote{http://opensoundcontrol.org/}
messages to a Python backend running on the same machine. The Python backend
processes the notes and may then send an OSC message containing notes to be
played to SuperCollider, which either generates the sound or forwards them to
an external MIDI controller for producing the sound. \autoref{fig:setup}
illustrates the main components of our system.

The SuperCollider component acts mostly as a bridge between the MIDI controllers
and the Python backend. It defines a set of handlers for routing MIDI input
messages to the backend via OSC messages, and handles OSC messages from the
backend. When a note on/off message is received from the backend, it can
either redirect to an external MIDI controller or produce the sound itself.
For the latter, the SuperCollider code loads a set of WAV files as well as a
few synthetic instruments for playback.

\section{Backend design}
\label{sec:backend}
At its core, the Python backend is running a continuous loop over a
customizable number of bars, each with a customizable number of beats. Each is
discretized it into 16th note segments (so one bar in 4/4 time signature will
have 16 intervals).  Multi-threading is used to allow for real-time response,
and we maintain a set of global variables that are shared across the different
threads, the most important of which are listed below:
\begin{itemize}
  \item $\mathbf{time\_signature}$: An object containing a pair of integers
    denoting the $numerator$ (4, 6, 7, etc.) and $denominator$ (4, 8, or 16) of
    the time signature. 
  \item $\mathbf{qpm}$: A float indicating the speed (quarters-per-minute) of
    playback. One quarter note is equal to four 16th notes, so this value indicates
    the time needed to process four 16th note events.
  \item $\mathbf{playable\_notes}$: A SortedList where we store each playable note
    event. Each element contains the type of playback event (click track, bass,
    drums, etc.), the note pitch, the instrument itself (bass, keyboard, hi-hat,
    bass drum, crash, etc.), and the 16th note in the bar where the event
    occurs.
  \item $\mathbf{bass\_line}$: Similar to $playable\_notes$ but containing only
    the current bassline.
  \item $\mathbf{accumulated\_primer\_melody}$: A list which will accumulate the
    note pitches played by the human improviser. Once enough notes have been
    accumulated they will be sent as a `primer' melody to MelodyRNN. This is
    discussed in more detail in the Improvisation section.
  \item $\mathbf{generated\_melody}$: A list containing the note pitches
    produced by MelodyRNN. When full, the note pitches played by the human will
    be replaced by the pitches in this buffer.
\end{itemize}

Our open source-code can be accessed at https://github.com/psc-g/Psc2.

\subsection{Click-track generation}
The first step is setting the number of bars, time signature, and tempo (qpm).
The user may change the number of bars, time signature numerator, and time
signature denominator via a set of buttons on the Nanokontrol2. The qpm may be
adjusted via a knob or by tapping the beat on a button. These define the length
and structure of the {\em sequence}, which the system will loop over.  Once
these are set the user may start playback by hitting the `play' button on the
Nanokontrol2.  This will start a click-track which will make use of 3 different
click sounds:
\begin{enumerate}
  \item The first will play on the first beat of the first bar, to indicate the
    start of the sequence. This is important for the user to known the start
    of the sequence when recording a bassline or chords.
  \item The second will play on the first beat of the remaining bars in the
    sequence (if at least two bars were selected)
  \item The third will play within each bar at a frequency marked by the time
    signature denominator: if the denominator is 4, it will play a click every 
    four 16th notes; if it is 8, it will play every two 16th notes; if it is 16
    it will play a click every 16th note.
\end{enumerate}

Once the click-track has been started, the user can place the system in one of
four {\em modes} via buttons on the Nanokontrol2. When SuperCollider is in
charge of producing sounds, ach mode uses a different instrument for playback.
\begin{itemize}
  \item {\bf bass:} The user can record a bassline which will be looped over.
    After entering this mode, recording begins as soon as a note is pressed and
    proceeds until the end of the sequence is reached.
  \item {\bf chords:} The user can play a set of chords to include in the loop
    playback. As in bass mode, recording begins as soon as a note is pressed
    and proceeds until the end of the sequence is reached.
  \item {\bf improv:} Used for improvising over the loop playback in a
    call-and-response between the human and the machine learning model. This
    mechanism is discussed in more detail in the Improvisation section.
  \item {\bf free:} Free-play mode,  where the human can improvise freely
    over the loop playback.
\end{itemize}

\subsection{Drums generation}
\begin{figure}[h]
  \centering
  \includegraphics[width=0.46\textwidth]{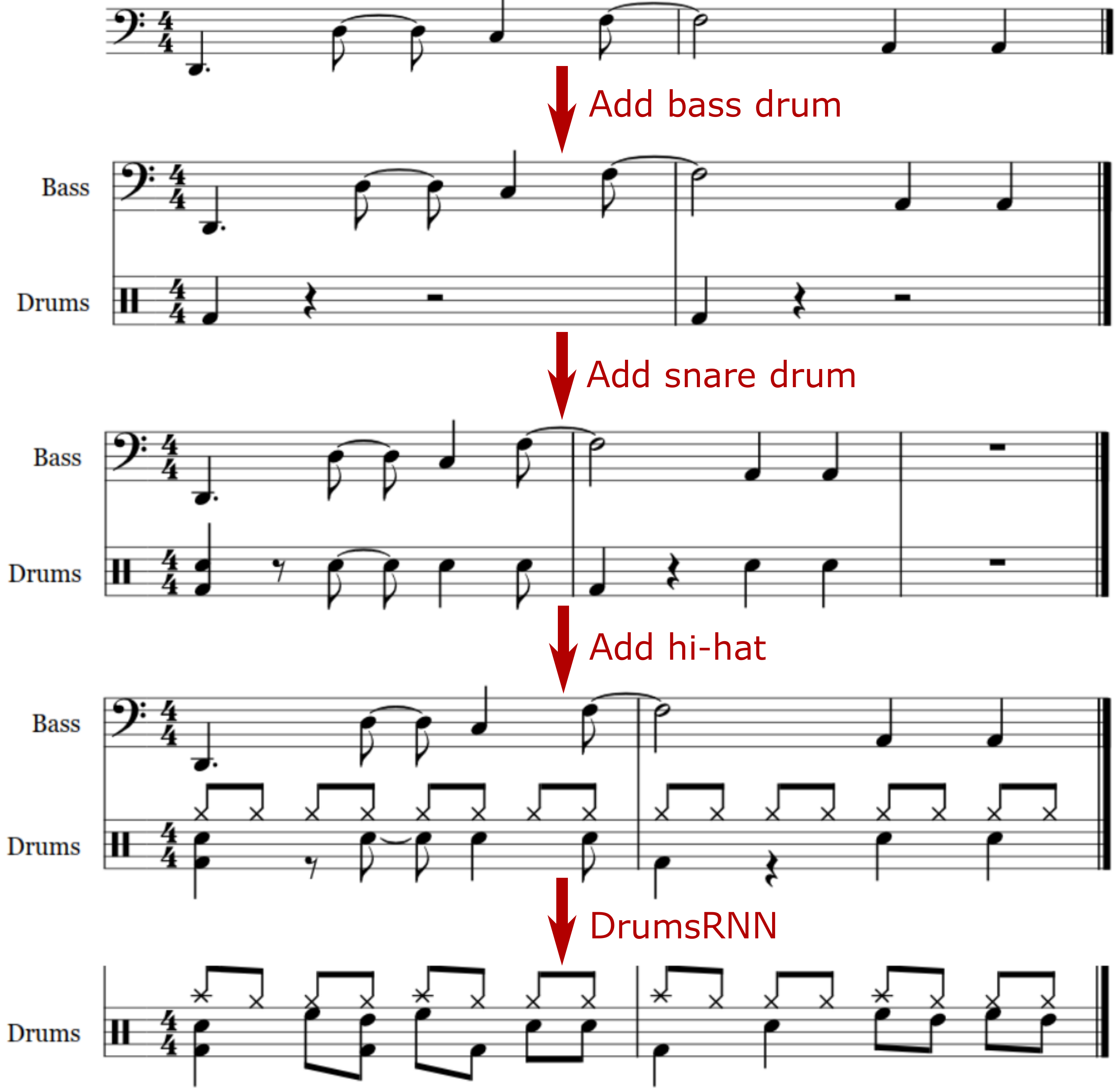}
  \caption{Building the drum beats. From top-to bottom: starting from a
  specified bassline, bass drum notes are added on the first beat
  of each bar, snare drum notes are added for each bass-note onset,
  and finally hi-hat notes are added at each 8th note. This deterministic
  drum beat can then be sent as a `primer' to DrumsRNN which will generate
  a new beat.}
  \label{fig:build_drums}
\end{figure}

Our system generates two types of drum beats: a deterministic one and and
another which is generated by a machine learning model. The deterministic
one is built off of the bassline as follows:
\begin{enumerate}
  \item A bass drum note is added at the first beat of every bar.
  \item A snare note is added at each bass note onset.
  \item Hi-hat notes are added at each 8th note (so every two 16th notes).
\end{enumerate}
By pressing one of the Nanokontrol2 buttons, this deterministic drum beat is
fed into DrumsRNN as a `primer' to produce a new beat.
\autoref{fig:build_drums} illustrates this process in musical notation.

\begin{figure*}[h]
  \centering
  \includegraphics[width=0.7\textwidth]{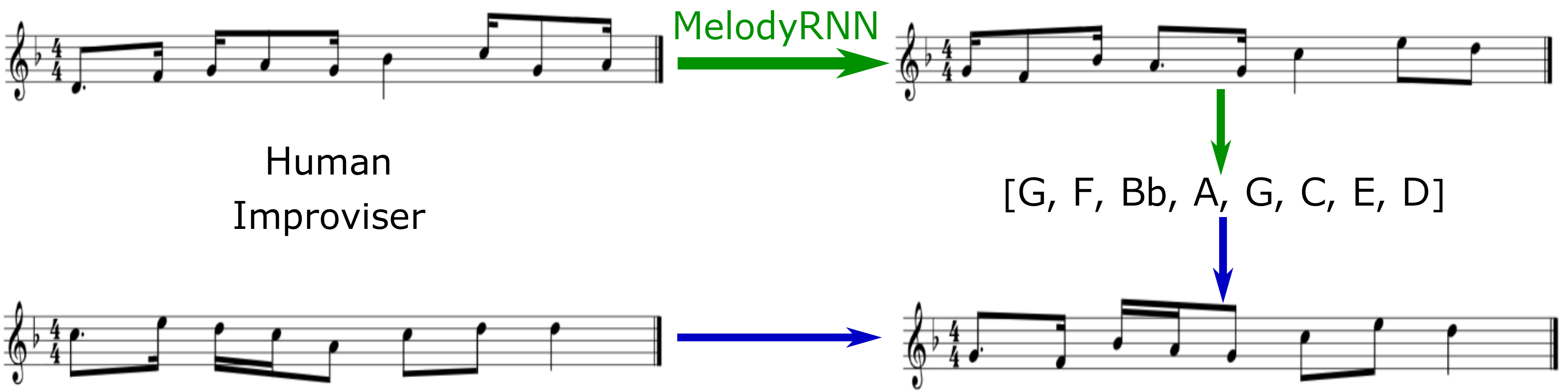}
  \caption{Building the hybrid improvisation. 1. The melody from the human
  improviser (top-left) is fed into MelodyRNN. 2. MelodyRNN produces a new
  melody (top-right). 3. The human improviser plays a new melody (bottom-left).
  4. A new hybrid melody is created by combining the pitches from the MelodyRNN
  model with the rhythm from the most recent human improvisation
  (bottom-right).}
  \label{fig:mlsplainer}
\end{figure*}

\subsection{Improvisation}
\label{sec:improv}
The improvisational part of our system is inspired on the call-and-response
improvisations that are common in traditional jazz. In these sections two or
more musicians take turns improvising over the same piece, and each musician
usually incorporates melodies and/or rhythms played by previous musicians into
their improvisations.

There are two main components to an improvisation: the pitches chosen and
the rhythm of the notes. In our experience playing with generative models,
such as MelodyRNN, we found that the rhythm of the melodies produced is not
very reflective of the types of rhythms observed from professional improvisers.
This may be due in large part to the 16th note quantization that is necessary
for training the models. To overcome this issue, we propose a hybrid approach:
the machine learning models provide the pitches, while the human provides
the rhythm.

The way this is achieved is as follows:
\begin{enumerate}
  \item Collect the pitches played by the human improviser in the
    $accumulated\_primer\_melody$ global buffer.
  \item Once the number of notes in the buffer is above a pre-specified
    threshold, the buffer is fed into MelodyRNN as a primer melody in a
    separate thread.
  \item When the MelodyRNN thread has finished generating a new melody, it
    will store {\em only the pitches} in the $generated\_melody$ buffer
    (the rhythmic information is dropped).
  \item When the main looper thread detects that the $generated\_melody$
    buffer has been filled, it will {\bf intercept} incoming notes played
    by the user and replace their {\em pitches} with the pitches stored
    in $generated\_melody$ (and removing said pitch from the buffer).
    \autoref{fig:mlsplainer} illustrates this process.
  \item Once $generated\_melody$ is empty, return to step 1.
\end{enumerate}

Our hybrid approach to machine-learning based improvisation allows us to
mitigate the two problems mentioned in the introduction: real-time
compatibility and stylistic personalization. The former is handled by
performing the inference in a separate thread and only using it when
it is available. The latter is handled by maintaining the rhythmic
inputs from the human performer. It has been found that rhythm can
significantly aid in facilitating melody detection \citep{jones87dynamic},
which we believe also carries over to enhancing the personalized
style of performance. Further, by leveraging the human's
rhythmic input, we are able to avoid having the limitation of the
16th-note quantization that the RNN models require.

We provide some videos demonstrating the use of this system at
https://github.com/psc-g/Psc2/tree/master/research/nips2018.

\section{Evaluation}
\label{sec:evaluation}
I have used this system for live jazz performance in a piano-drums
duet. The songs performed were some that I regularly perform with my trio
(www.psctrio.com), and the system was engaged during the improvisation sections
of these songs. Since there was a human drummer performing, only the
improvisation (MelodyRNN) part of our system was used. Some of my thoughts
from these performances are listed below.

\subsection{Strengths}
\begin{itemize}
  \item The system was able to respond in real-time: there was no noticeable
    lag between playing a note on the piano and hearing the resulting sound.
  \item Since the system produces MIDI outputs, I could use any desired sound
    from my analogue synthesizer, which made for a more organic sound.
  \item The audience (many of which are familiarized with my style)
    reported not noticing that there was an external system affecting the
    improvisations (they were only made aware of it after the show). I take
    this as a very successful signal, as it demonstrates that this approach is
    able to maintain the key stylystic elements of the human performer.
  \item I did find that I had to adjust my improvisations to account
    for the element of uncertainty that the system introduced. Specifically,
    not knowing which pitches would be produced from they key-presses forced
    me away from my go-to lines and into a new creative space.
\end{itemize}

\subsection{Weaknesses}
\begin{itemize}
  \item I found it difficult to use the system in songs with many
    harmonic shifts. Since we did not impose boundaries for recording the
    human performer's notes (such as at the start and end of a jazz chorus),
    the times when the system would begin replacing the human's notes were
    somewhat arbitrary, possibly resulting in melodies that are not properly
    aligned with the underlying harmony. On {\em modal} songs (where the
    harmony does not shift) I had no issues.
  \item Sometimes the system would come in at the ``wrong'' time; specifically,
    when I was in the middle of a line that was cut short, resulting in an
    awkard melody.
  \item If I was at one register of the piano when the system began
    recording my notes and ended up in a different register right before the
    system engaged, when the MelodyRNN notes began replacing mine, they
    would tend to start in the first register, causing a big jump in pitch. This
    is generally discouraged from a musical perspective, which forced me
    to improvise mostly within two octaves.
\end{itemize}

\section{Conclusion and Future Work}
\label{sec:conclusion}
In this paper we have introduced a system that enables the integration
of out-of-the-box deep learning models for live improvisation. We have
designed it in a way that it does not require machine learning expertise
to use, and can be extended to other types of musical generative models
with little effort. Our hybrid approach for generating machine-learning
based improvisations maintains the style of the human improviser while
producing novel improvised melodies.

Although our system was built with MelodyRNN and DrumsRNN, the
setup can be used with any musical generative model with relatively little
effort. Along these lines, one avenue we would like to explore in the future is
the incorporation of models which do not require quantization, such as
PerformanceRNN \citep{simon17performance}; one challenge is to ensure
that the personal style of the human improviser is maintained.

Expert musicians are able to produce high-quality improvisations {\em
consistently} from having honed their craft over many years of practice. A
common frustration with these artists, however, is that they often find their
improvisations too predictable, and struggle escaping their ``pocket''.  Our
hope is that systems like the one we are proposing here can push expert
musicians, and artists in general, out of their comfort zone and in
new directions they may not have chosen to go to on their own. The experience
of the pianist we reported in the previous section perfectly showcases
this.

We hope to improve the system by allowing the performer to have more control
over when the system begins recording, and when the system replaces the notes.
We have already begun experimenting with this extra modality using a MIDI
footpedal. Initial experiments suggest this added level of control mitigates
for many of the issues raised by the human performer regarding the timing
of when the system is engaged, while maintaining the novelty of the melodies
produced.

%






\bibliographystyle{iccc}
\bibliography{jazz_improv}

\end{document}